\documentclass [prl,twocolumn,superscriptaddress,preprintnumbers]{revtex4}

\usepackage {bm}
\usepackage {graphicx}

\begin{document}

\title{Femtosecond Diffractive Imaging with a Soft-X-ray Free-Electron Laser}

\author{Henry N. Chapman}
\email{henry.chapman@llnl.gov}
\affiliation{
University of California, Lawrence Livermore National Laboratory,
7000 East Avenue, Livermore CA 94550, USA.}
\affiliation{
Center for Biophotonics Science and Technology, University of
California, Davis, 2700 Stockton Blvd., Suite 1400, Sacramento, CA
95817, USA.}

\author{Anton Barty}
\affiliation{
University of California, Lawrence Livermore National Laboratory,
7000 East Avenue, Livermore CA 94550, USA.}

\author{Michael J. Bogan}
\affiliation{
University of California, Lawrence Livermore National Laboratory,
7000 East Avenue, Livermore CA 94550, USA.}

\author{S\'ebastien Boutet}
\affiliation{
University of California, Lawrence Livermore National Laboratory,
7000 East Avenue, Livermore CA 94550, USA.}
\affiliation{
Stanford Synchrotron Radiation Laboratory, Stanford Linear
Accelerator Center, 2575 Sand Hill Road, Menlo Park, California 94305,
USA.}
\affiliation{
 Laboratory of Molecular Biophysics, Department of Cell and
Molecular Biology, Uppsala University, Husargatan 3, Box 596, SE-75124
Uppsala, Sweden.}

\author{Matthias Frank}
\affiliation{
University of California, Lawrence Livermore National Laboratory,
7000 East Avenue, Livermore CA 94550, USA.}

\author{Stefan P. Hau-Riege} 
\affiliation{
University of California, Lawrence Livermore National Laboratory,
7000 East Avenue, Livermore CA 94550, USA.}

\author{Stefano Marchesini}
\affiliation{
University of California, Lawrence Livermore National Laboratory,
7000 East Avenue, Livermore CA 94550, USA.}

\affiliation{
Center for Biophotonics Science and Technology, University of
California, Davis, 2700 Stockton Blvd., Suite 1400, Sacramento, CA
95817, USA.}

\author{Bruce W. Woods}
\affiliation{
University of California, Lawrence Livermore National Laboratory,
7000 East Avenue, Livermore CA 94550, USA.}

\author{Sa\u{s}a Bajt}
\affiliation{
University of California, Lawrence Livermore National Laboratory,
7000 East Avenue, Livermore CA 94550, USA.}

\author{W. Henry Benner}
\affiliation{
University of California, Lawrence Livermore National Laboratory,
7000 East Avenue, Livermore CA 94550, USA.}

\author{Richard A. London}
\affiliation{
University of California, Lawrence Livermore National Laboratory,
7000 East Avenue, Livermore CA 94550, USA.}
\affiliation{
Center for Biophotonics Science and Technology, University of
California, Davis, 2700 Stockton Blvd., Suite 1400, Sacramento, CA
95817, USA.}

\author{Elke Pl\"onjes}
\author{Marion Kuhlmann} 
\author{Rolf Treusch}
\author{Stefan D\"usterer}
\author{Thomas Tschentscher} 
\author{Jochen R. Schneider} 
\affiliation{
Deutsches Elektronen-Synchrotron, DESY, Notkestra\ss e 85, D-22607
Hamburg, Germany.}

\author{Eberhard Spiller}
\affiliation{Spiller X-ray Optics, Livermore CA 94550, USA.}

\author{Thomas M\"oller}
\author{Christoph Bostedt} 
\author{Matthias Hoener}
\affiliation{
Institut f\"ur Atomare Physik, Technische Universit\"at Berlin,
Hardenbergstra\ss e 36, PN 3-1, 10623 Berlin, Germany.}

\author{David A. Shapiro} 
\affiliation{
Center for Biophotonics Science and Technology, University of
California, Davis, 2700 Stockton Blvd., Suite 1400, Sacramento, CA
95817, USA.}

\author{Keith O. Hodgson}
 \affiliation{
Stanford Synchrotron Radiation Laboratory, Stanford Linear
Accelerator Center, 2575 Sand Hill Road, Menlo Park, California 94305,
USA.}

\author{David van der Spoel} 
\author{Florian Burmeister} 
\author{Magnus Bergh}
\author{Carl Caleman} 
\author{G\"osta Huldt} 
\author{M. Marvin Seibert} 
\author{Filipe R.N.C. Maia} 
\affiliation{
 Laboratory of Molecular Biophysics, Department of Cell and
Molecular Biology, Uppsala University, Husargatan 3, Box 596, SE-75124
Uppsala, Sweden.}

\author{Richard W. Lee}
\author{Abraham Sz\"oke} 
\affiliation{
University of California, Lawrence Livermore National Laboratory,
7000 East Avenue, Livermore CA 94550, USA.}
\affiliation{
 Laboratory of Molecular Biophysics, Department of Cell and
Molecular Biology, Uppsala University, Husargatan 3, Box 596, SE-75124
Uppsala, Sweden.}

\author{Nicusor Timneanu}
\affiliation{
 Laboratory of Molecular Biophysics, Department of Cell and
Molecular Biology, Uppsala University, Husargatan 3, Box 596, SE-75124
Uppsala, Sweden.}

\author{Janos Hajdu}
\email{janos.hajdu@xray.bmc.uu.se}
 \affiliation{
Stanford Synchrotron Radiation Laboratory, Stanford Linear
Accelerator Center, 2575 Sand Hill Road, Menlo Park, California 94305,
USA.}
\affiliation{
 Laboratory of Molecular Biophysics, Department of Cell and
Molecular Biology, Uppsala University, Husargatan 3, Box 596, SE-75124
Uppsala, Sweden.}

\begin{abstract}
Theory predicts \cite{r1,r2,r3,r4}
that with an ultrashort and
extremely bright coherent X-ray pulse, a single diffraction pattern
may be recorded from a large macromolecule, a virus, or a cell before
the sample explodes and turns into a plasma. Here we report the first
experimental demonstration of this principle using the FLASH soft
X-ray free-electron laser. An intense 25 fs, $4\times 10^{13}$ W/cm$^2$ pulse,
containing $10^{12}$ photons at 32 nm wavelength, produced a coherent
diffraction pattern from a nano-structured non-periodic object, before
destroying it at 60,000 $^\circ$K. A novel X-ray camera assured single photon
detection sensitivity by filtering out parasitic scattering and plasma
radiation. The reconstructed image, obtained directly from the
coherent pattern by phase retrieval through oversampling
\cite{r5,r6,r7,r8,r9}, shows no measurable damage, and extends to
diffraction-limited resolution. A three-dimensional data set may be
assembled from such images when copies of a reproducible sample are
exposed to the beam one by one \cite{r10}.
\end{abstract}

\preprint{UCRL-JRNL-219848}

\maketitle
X-ray free-electron lasers (X-ray FELs) are expected to permit
diffractive imaging at high-resolutions of nanometer- to
micrometer-sized objects without the need for crystalline periodicity
in the sample \cite{r1,r2,r3,r4}. High-resolution structural studies
within this size domain are particularly important in materials
science, biology, and medicine. Radiation-induced damage and sample
movement prevents the accumulation of high-resolution scattering
signals for such samples in conventional
experiments \cite{r11,r12}. Damage is caused by energy deposited into the
sample by the very probes used for imaging, e.g. photons, electrons,
or neutrons. At X-ray frequencies inner shell processes dominate the
ionisation of the sample; photoemission is followed by Auger or
fluorescence emission and shake excitations. The energies of the
ejected photoelectrons, Auger electrons, and shake electrons differ
from each other, and these electrons are released at different times,
but within about ten femtoseconds, following photoabsorption
\cite{r1,r13}. Thermalisation of the ejected electrons through
collisional electron cascades is completed within 10-100 femtoseconds
\cite{r14,r15}. Heat transport, diffusion and radical reactions take
place over some picoseconds to milliseconds.

Radiation tolerance in the X-ray beam could be substantially extended,
if we could collect diffraction data faster than the relevant damage
processes \cite{r1,r16}. This approach requires very short and very
bright X-ray pulses, such as those expected from short-wavelength
free-electron lasers. However, the large amount of energy deposited
into the sample by a focused FEL pulse will ultimately turn the sample
into a plasma. The question is when exactly would this happen? There
are no experiments with X-rays in the relevant time and intensity
domains, and our current understanding of photon-material interactions
on ultra-short time scales and at high X-ray intensities is,
therefore, limited. Computer simulations based on four different
models \cite{r1,r2,r3,r4} postulate that a near-atomic resolution
structure could be obtained by judicious choice of pulse length,
intensity and X-ray wavelength, before the sample is stripped from its
electrons and is destroyed in a Coulomb explosion. Near-atomic
resolution imaging with X-ray FEL pulses faces other formidable
challenges that must be addressed, such as developing the ability to
record low-noise and interpretable diffraction data under the extreme
illumination conditions expected from a focused FEL pulse.

Our experimental demonstration of ``flash diffractive imaging" uses
the first soft X-ray FEL in the world, the FLASH facility (formerly
known as the VUV-FEL) at the Deutsches Elektronen-Synchrotron (DESY)
in Hamburg \cite{r17}.  FLASH generates high power soft X-ray pulses
by the principle of self-amplification of spontaneous emission (SASE)
\cite{r18}: a relativistic electron pulse from a superconducting linear
accelerator makes a single pass through a periodic magnetic field of
an undulator. During the high-gain lasing process, the electrons,
perturbed by the magnetic field of the undulator and by their own
photon field, form coherent micro-bunches, which behave like a single
giant charge, producing strong amplification. For our experiment,
FLASH was operating in an ultrashort pulse mode \cite{r17}, resulting
in 25 fs coherent FEL pulses with about $10^{12}$ photons in a pulse.

\begin{figure}[tbp]
\centerline{\includegraphics[width=0.45\textwidth]{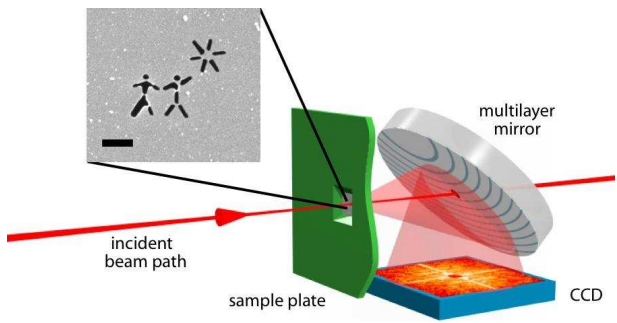}}
\caption{
Schematic diagram of the experimental apparatus. The FEL beam is
incident from the left and is focused to a 20-$\mathrm{\mu}$m spot on the
sample, which is a 20-nm thick transmissive silicon nitride membrane
with a picture milled through its entire thickness using a FIB (this
is enlarged in the inset, and the scale bar indicates 1 $\mu$m).  The
direct beam passes through the sample window and exits the camera
through a hole in a graded multilayer planar mirror. The diffracted
light from the sample reflects from that mirror onto a CCD
detector. The contour lines on the mirror depict lines of constant
incidence angle (constant multilayer period).  The on-axis path length
from the sample to the detector is 55 mm. For 32 nm radiation and
objects smaller than 20 $\mu$m, this distance is in the far field,
where the diffraction pattern is equal to the Fourier transform of the
exit wave \cite{r27}. The numerical aperture of the detector is 0.25.}
\end{figure}

Figure 1 shows our experimental arrangement. Diffractive imaging is
elegant in its simplicity: a coherent X-ray beam illuminates the
sample, and the far-field diffraction pattern of the object is
recorded on an area detector. We focused a coherent 25 fs X-ray pulse
from FLASH to achieve a peak intensity of $(4\pm 2)\times 10^{13}$
$\mathrm{W/cm^2}$ on the sample.  We recorded the far-field diffraction
pattern of the object on a novel detector centred on the forward
direction (see Methods). The image information encoded in the coherent
diffraction pattern is similar to a hologram \cite{r19}, except that the object
acts as its own scattering reference. Image reconstruction was
performed by phase retrieval using our iterative transform algorithm,
Shrinkwrap \cite{r8} (see Methods). Unlike similar algorithms
\cite{r7,r20,r21,r22,r23},
Shrinkwrap solves the phase problem without requiring any a priori
knowledge about the object.

\begin{figure}[tbp]
\centerline{\includegraphics[width=0.35\textwidth]{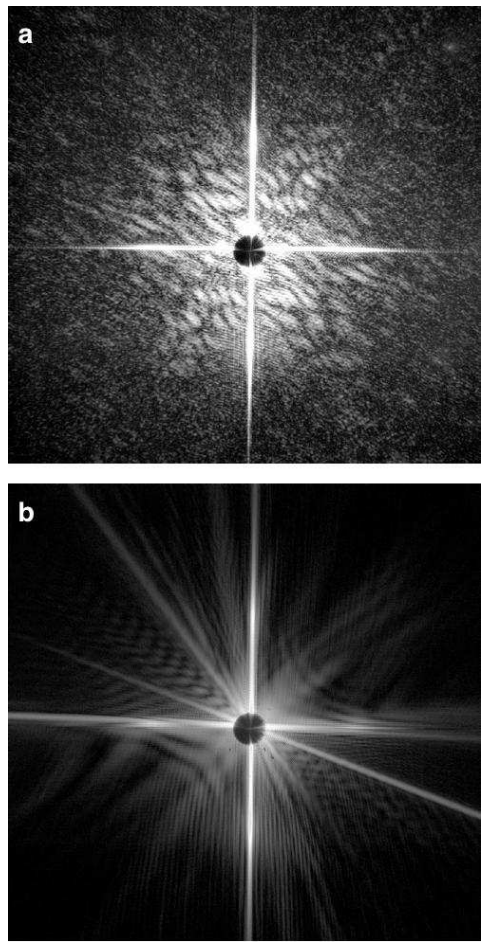}}
\caption{
 Flash X-ray coherent diffraction patterns. (a) Coherent
diffraction pattern recorded for a single 
$(4\pm 2)\times 10^{14}$ W/cm$^2$, 
$25\pm 5$ fs
pulse, and (b) for the subsequent pulse of similar intensity and
duration, 20 s later, showing diffraction from the damage caused by
the pulse that formed (a). The intensity is shown on a logarithmic
grey scale with black denoting 1 photon/pixel and white denoting 2000
photons/pixel for (a) and 50,000 photons/pixel for (b). The entire
patterns are shown as detected by the CCD, and extend to a diffraction
angle of 15$^\circ$ at the midpoint of the edges (corresponding to a momentum
transfer of 8.1 $\mu$m$^{-1}$).}
\end{figure}

The ultrafast
coherent diffraction pattern of a nano-structured non-periodic object
is shown in Fig. 2(a). The object was a micron-sized pattern cut
through a partially-transparent silicon nitride membrane with a
focused-ion beam (FIB), and it is shown in the insert of Figure 1. The
pattern extends to a diffraction angle of $15^\circ$ at the midpoint of its
edge. Based on low-fluence optical parameters \cite{r24}, we estimate 
\cite{r3,r25} that the absorbed energy density was approximately 20 eV/atom 
in the silicon nitride and that the sample reached a temperature of about
60,000 $^\circ$K before vaporizing. A second diffraction pattern taken 20 s
after the first exposure is shown in Figure 2(b). This shows
diffraction from a hole left in the membrane caused by the first
pulse. That is, the first pulse utterly destroyed the sample but not
before a diffraction pattern of the apparently undamaged object could
be recorded. Images of the object obtained with a scanning electron
microscope (SEM), before and after FEL exposure, are shown in Figure
3.

\begin{figure*}[htbp]
\centerline{\includegraphics[width=0.95\textwidth]{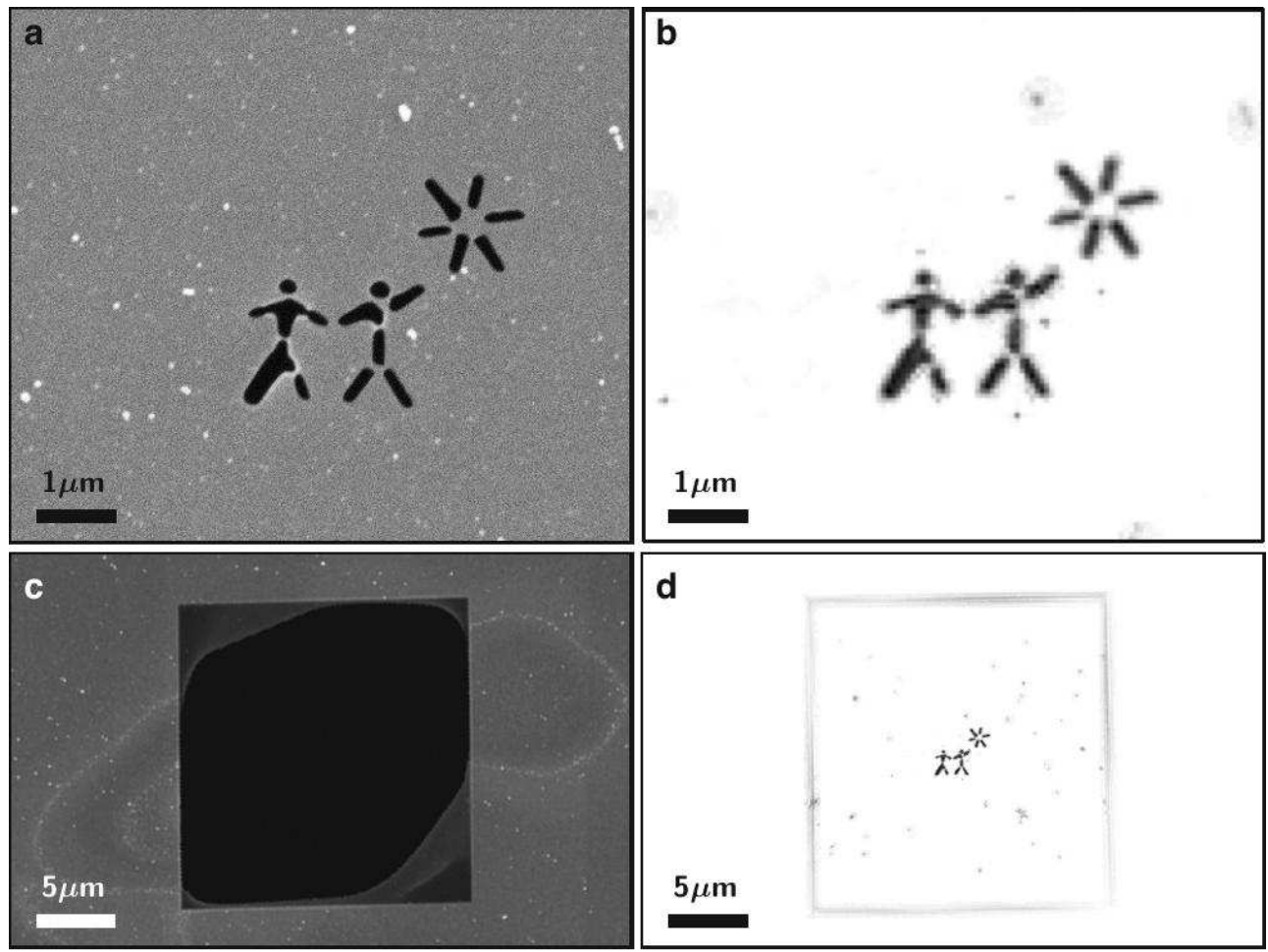}}
\caption{
 The reconstructed X-ray image shows no evidence of the
damage caused by the pulse. (a) Scanning electron microscope (SEM)
image of the sample before exposure to the FEL beam. The 20 nm thick
sample was held in a square supporting window that is 20 $\mu$m wide. (b)
and (d) Image reconstructed, from the ultrafast coherent diffraction
pattern of Fig. 2 (a), by phase retrieval and inversion using the
Shrinkwrap algorithm \cite{r8}. The squared modulus of the retrieved complex
image is displayed on a linear grey scale from zero scattered
photons/pixel (white) to $1.5\times 10^6$ scattered photons/pixel
(black). Pixel size in the reconstruction = 62 nm in (b),
corresponding to the half period of the finest spatial frequency that
can be recorded on our camera at 32 nm wavelength. The retrieved image
clearly shows the silicon window edge (in d), the FIB pattern, and
dirt particles. (c) SEM image of the test sample after the exposures
to the FEL beam, showing the square 20-$\mu$m window and some remaining
silicon nitride, as well as visible damage to the silicon support
caused by the non-circular beam. The scale bar for (a) and (b) is 1 $\mu$m
and the scale bar for (c) and (d) is 5 $\mu$m.}
\end{figure*}

The main features of the diffraction pattern of Figure 2(a) are
speckles and strong vertical and horizontal lines that pass through
the centre of the pattern. The horizontal and vertical lines are
caused by interference of the waves diffracting from the opposite
edges of the square window frame that holds the silicon nitride
membrane. The speckles correspond to two length scales of the
sample. The modulations of $\sim 60$ pixels (measured diagonally) in the
diffraction pattern near the centre correspond to the narrow 2.5
$\mu$m diagonal dimension of the object; and the finer speckles of
about 16 pixels correspond to the distance between the picture object
and the window frame in which it is centered. The speckles remain well
defined out to the edge of the detector, although their visibility
diminishes with scattering angle. This may be due to the fact that at
the high diffraction angles at the edge of the CCD detector, the
optical path difference between rays diffracting from points in the
object transversely separated by 20 $\mu$m (the sample window size) is
$\sin\left ( 15^\circ\right ) \times 20$ $\mu\mathrm{m}= 5\mu$m. 
This is comparable to the
length of a 25 fs pulse, which is 7.5 $\mathrm{\mu}$m. That is, the overlap of the
beams in time (and hence interference between them) only occurs for
one third of the pulse at high angles.

Figures 3(b) and (d) show the image of the object reconstructed
directly from the diffraction pattern of Figure 2(a). The angular
acceptance $\alpha$, of our detector is $15^\circ$ at the midpoint of
the detector edges, and $20^\circ$ at the corners. The diffraction limited
resolution length is $\lambda/(2 \sin \alpha ) = 62$ nm 
for a wavelength of $\lambda= 32$ nm. 
This length is defined as the half-period of the finest spatial
frequency in the image, equal to an image pixel width. Along diagonal
directions the increased CCD acceptance gives a resolution length of
43 nm. The actual image resolution would be worse than the diffraction
limit if the retrieved phases were incorrect, in the same way that
phase errors in a lens cause image aberrations.  We estimate the image
resolution by computing the Phase-Retrieval Transfer Function
(PRTF) \cite{r9,r23}, shown in Figure 4.  This function represents the
confidence for which the diffraction phases have been retrieved and is
calculated by comparing the Fourier amplitudes of the average of
multiple independent reconstructions to the measured diffraction
amplitudes.  Where the phase of a particular Fourier component is
consistently retrieved the complex values add in phase, whereas if the
phase is random the sum will approach zero.  The PRTF is thus equal to
unity when the phase is consistently retrieved and zero when the phase
is unknown.  We use the convention that the resolution is given by the
point where the PRTF drops to 1/e (reference \cite{r23}), which for this image
occurs at the resolution limit (62 nm).  We note that the same
experimental geometry deployed on a hard X-ray free electron laser
operating at 0.15 nm wavelength would yield a diffraction-limited
resolution length of 0.3 nm.  

\begin{figure}[htbp]
\centerline{\includegraphics[width=0.45\textwidth]{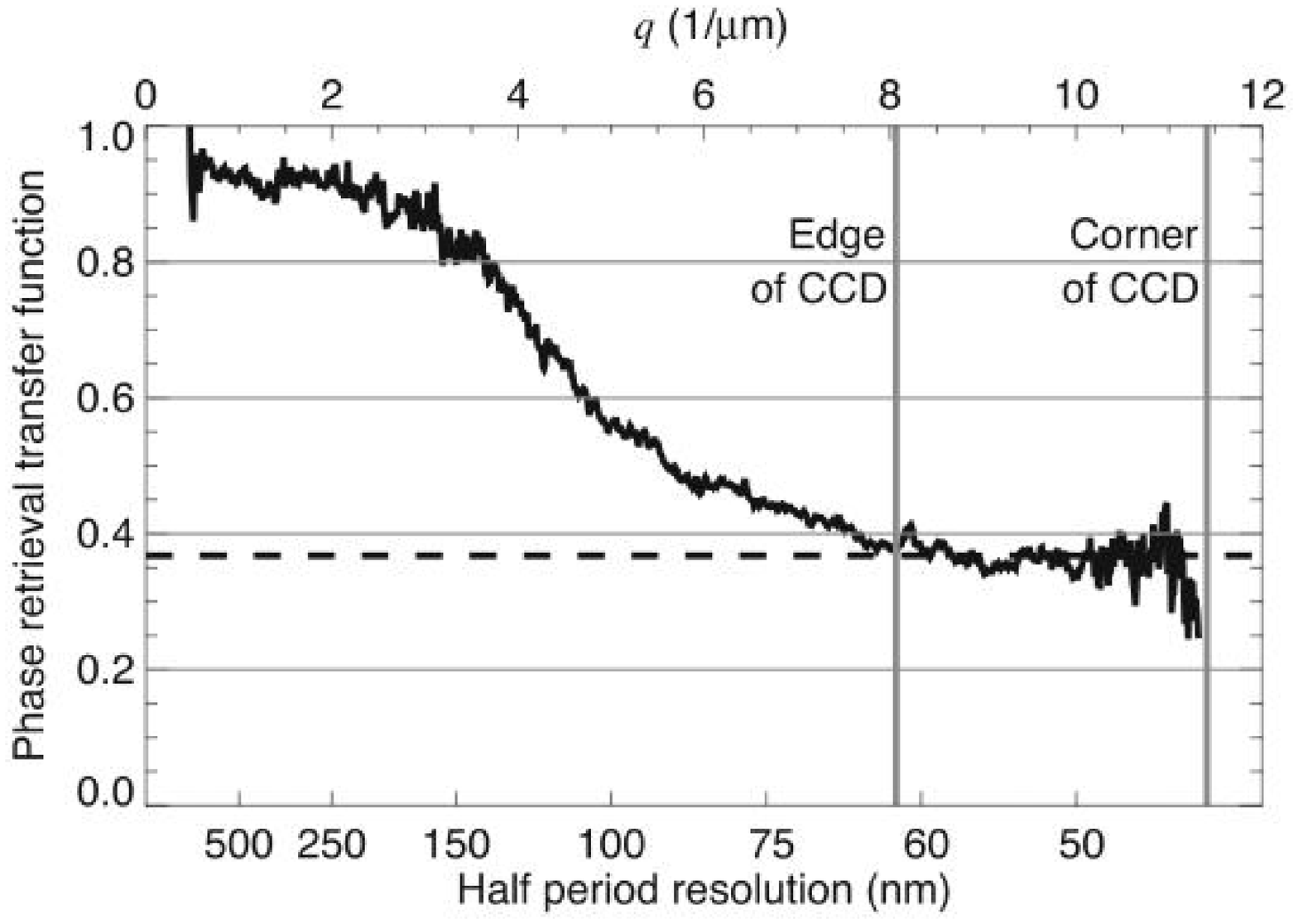}}
\caption{
The image is reconstructed to the diffraction
limit. Phase-retrieval transfer function (PRTF) \cite{r9,r23} for the
reconstructed image shown in Figure 3(b) and (d), averaged over shells
of constant momentum transfer where is the wavelength and the
diffracted angle. The PRTF is equal to unity when the phase is
consistently retrieved and zero when the phase is unknown. Using the
convention \cite{r23} that the resolution is given by the point where the PRTF
drops to 1/e, the resolution of our reconstruction is estimated to be
62 nm.}
\end{figure}

The ``lensless'' imaging method used here
can be extended to atomic resolution, which will require shorter
wavelength X-rays and tighter focusing than demonstrated here.  Hard
X-ray FELs are currently being developed that will create pulses
which, when focused on the sample, will produce five orders of
magnitude higher photon intensities than used here. An understanding
of photon-material interactions on ultra-short time scales and at high
x-ray intensities is fundamentally important to all experiments with
X-ray lasers. This area of science is virtually unexplored. The FLASH
free-electron laser in Hamburg is the first radiation source to permit
experiments near the relevant photon energies and intensities. Our
present results validate the concept of single-shot imaging with
extremely intense and ultra short soft X-ray pulses that are capable
of destroying anything in their path. The resulting diffraction
pattern carries high-resolution structural information about the
object, and the resolution of the reconstructed image extends to the
diffraction limit. This indicates that significant damage occurs only
after the ultrashort 25 fs FEL pulse traverses the sample. These
results have implications for studying non-periodic molecular
structures in biology, or in any other area of science and technology
where structural information with high spatial and temporal resolution
is valuable. They also point to the viability of nanometer- to
atomic-resolution imaging of non-periodic and non-crystalline
objects \cite{r1,r2,r3,r4} with hard X-ray FELs.

\section{Methods}

Samples consisted of a 20-nm thick silicon nitride membrane spanning a
20 $\mu$m wide square silicon window. The pattern was cut through the
membrane with a dual-beam focused ion beam instrument (FEI, National
Center for Electron Microscopy, Lawrence Berkeley National
Laboratory), using a 500 pA beam of 30 keV $\mathrm{Ga^+}$ ions.  The
20 nm thick silicon nitride membrane has a transmission of 44\% at a
wavelength of 32 nm, and causes an estimated phase advance of 20$^\circ$
relative to propagation through the same length of vacuum, calculated
from the known low-fluence optical constants \cite{r24}.  

Experiments were
performed in vacuo as everything in the direct beam contributes to the
diffraction pattern. The samples were placed in a vacuum vessel 70 m
from the FEL undulator. The FEL pulse was focused to a 30 $\mu$m$\times$20
$\mu$m focal spot on the sample with a 2-m focal length ellipsoidal
mirror in the beam line. Single pulses were selected with a fast
shutter. Due to the source coherence, no aperture was needed to select
a coherent patch of the beam, nor was a monochromator used to select a
narrow wavelength band of the radiation.

A novel X-ray camera was developed to record low-noise diffraction
data from the sample in the forward direction (see Figure 1).  In this
camera, a graded multilayer plane mirror separates the diffracted beam
from the direct beam, and the intense direct beam passes harmlessly
through the hole in the centre of the mirror without damaging the
detector. The diffracted light reflects onto a back-illuminated
direct-detection CCD chip (Princeton Instruments in-vacuum PI-MTE
CCD), containing $1300\times 1340$ square pixels of 20 $\mu$m
width. The resonant X-ray multilayer of the planar mirror consists of
layers of Si, Mo, and $\mathrm{B_4C}$, and was fabricated so that the
layer period varies from 18 nm to 32 nm across the mirror. The
variation in multilayer period matches the variation in the angle of
incidence of rays emanating from the sample and which strike the
mirror.  This angle varies from $30^\circ$ to $60^\circ$, as depicted
by the contour lines on the mirror in Figure 1. The gradient was
achieved by sputter-depositing the multilayer materials through a mask
onto the rotating substrate, so that the time-averaged deposition gave
the desired material thickness at each point on the mirror.  The 32-nm
reflectivity across the mirror was 45\%, as measured at a
synchrotron-based reflectometer \cite{r26}.  Only X-rays within a
bandwidth of 9 nm and which propagate from near the sample interaction
point are efficiently reflected.  Broadband plasma emission from the
sample is filtered out by the resonant mirror.  Also, off-axis
radiation scattered from beamline components is reflected at less than
1\% and hence filtered from the diffraction pattern.  The reflectivity
of the coating diminishes smoothly to zero close to the edge of the
central hole, due to decoherence of the coating layers caused by the
underlying substrate roughness where the hole was cored.  This ``soft
edge" reduces scatter from the hole, whose shadow can be seen as a
dark circle at the centre of the patterns in Figure 2.  The on-axis
path length of the reflected beam from the sample to the CCD was 55
mm, and for 32 nm radiation and objects smaller than 20 $\mu$m, this
distance is in the far field, where the diffraction pattern is equal
to the Fourier transform of the exit wave \cite{r27}.  

Image reconstruction was carried out with the Shrinkwrap algorithm
\cite{r8}. Phase retrieval in Shrinkwrap is a non-linear optimization
problem in a high-dimensional phase space. The dimensionality is equal
to the number of phases to be retrieved: 1.7 million in this case. The
solution is obtained iteratively by sequentially enforcing known
constraints in diffraction and image spaces. We specifically aim for
diffraction phases that are such that the waves re-interfering to form
the image must all destructively cancel in areas outside the object's
boundary (called its support), and that the amplitudes of the discrete
Fourier transform of the image match the measured diffraction
amplitudes (which must be measured finely enough to include enough
empty space beyond the object to constrain the phases). Other
iterative transform algorithms usually require that the support of the
object be known a priori, and the closer the support to the actual
object boundary, the better the reconstruction. Shrinkwrap, however,
periodically refines the support constraint from the current estimate
of the image.  The support constraint is calculated every 70
iterations by selecting pixels with intensity values greater than 0.2
times the maximum image intensity, after first blurring the image with
a Gaussian kernel.  The blurring kernel is initially set to 3 pixels
full-width half-maximum (FWHM) and is gradually reduced to 0.7 pixels
FWHM by iteration 5000. The final support is that found four update
cycles prior to the point where the normalized image error \cite{r9}
exceeds a value of 0.2. This stopping criterion is typically reached
in 3000 to 4000 iterations.  During the iterations we did not
constrain the intensity or phase in the region in the mirror hole,
which contains the unrecorded zero spatial frequency, nor did we
constrain the object to be real or positive. We performed many
reconstructions, starting each time from random phases. Each
reconstructed image varied slightly due to the fact that with photon
shot noise there is no true solution that exactly satisfies all
constraint sets. However, each image determined from the final iterate
was clearly recognizable as compared with the SEM image. The image
estimate, displayed in Fig. 3 (b) and (d) is the average of 250
independent reconstructions.

\section{Acknowledgements}

Special thanks are due to the scientific and technical staff of FLASH
at DESY, Hamburg, in particular to J. Feldhaus, R. L. Johnson,
U. Hahn, T. Nu\~nez, K. Tiedtke, S. Toleikis, E. L. Saldin,
E. A. Schneidmiller, and M. V. Yurkov. We also thank R. Falcone,
M. Ahmed and T. Allison for discussions, J. Alameda, E. Gullikson,
F. Dollar, T. McCarville, F. Weber, J. Crawford, C. Stockton,
W. Moberlychan, M. Haro, A. Minor, H. Thomas and E. Eremina for
technical help with these experiments. This work was supported by the
following agencies: The U.S. Department of Energy (DOE) under Contract
W-7405-Eng-48 to the University of California, Lawrence Livermore
National Laboratory (the project 05-SI-003 was funded by the
Laboratory Directed Research and Development Program at LLNL); The
National Science Foundation Center for Biophotonics, University of
California, Davis, under Cooperative Agreement PHY 0120999; The
National Center for Electron Microscopy and the Advanced Light Source,
Lawrence Berkeley Lab, under DOE Contract DE-AC02-05CH11231; Natural
Sciences and Engineering Research Council of Canada (NSERC
Postdoctoral Fellowship to MB); the U.S. Department of Energy Office
of Science to the Stanford Linear Accelerator Center; the European
Union (TUIXS); The Swedish Research Council; The Swedish Foundation
for International Cooperation in Research and Higher Education; and
The Swedish Foundation for Strategic Research.  

\noindent
\textbf{Competing Financial Interests:} The authors declare that they have no competing financial
interests.

\end{document}